\documentclass[twocolumn,preprintnumbers,amsmath,amssymb,floatfix,superscriptaddress,nofootinbib]{revtex4}

\usepackage{amsmath,hyperref,amssymb}
\usepackage{graphicx}% Include figure files
\usepackage{dcolumn}% Align table columns on decimal point
\usepackage{bm}% bold math
\usepackage{verbatim}
\usepackage{tabularx}
\usepackage{slashed}
\usepackage{cancel}
\usepackage[boxsize=2em,aligntableaux=center]{ytableau}
\usepackage{float}
\usepackage{color}

\def\be{\begin{equation}}
\def\ee{\end{equation}}
\def\bea{\begin{eqnarray}}
\def\eea{\end{eqnarray}}

\definecolor{p}{RGB}{223,0,225}
\definecolor{o}{RGB}{255,94,1}
\definecolor{g}{RGB}{0,150,85}

\begin{document}

\vspace*{-30mm}

\title{Warming up cold inflation}

\author{William DeRocco}
\affiliation{Stanford Institute for Theoretical Physics, \\
Stanford University, Stanford, CA 94305, USA}
\author{Peter W. Graham}
\affiliation{Stanford Institute for Theoretical Physics and Kavli Institute for Particle Astrophysics \& Cosmology, Stanford University, Stanford, CA 94305, USA}
\author{Saarik Kalia}
\affiliation{Stanford Institute for Theoretical Physics, \\
Stanford University, Stanford, CA 94305, USA}

\vspace*{1cm}

\begin{abstract} 

The axion is a well-motivated candidate for the inflaton, as the radiative corrections that spoil many single-field models are avoided by virtue of its shift symmetry. However, axions generically couple to gauge sectors. 
As the axion slow-rolls during inflation, this coupling can cause the production of a non-diluting thermal bath, a situation known as ``warm inflation.'' This thermal bath can dramatically alter inflationary dynamics and observable predictions.
In this paper, we demonstrate that a thermal bath can form for a wide variety of initial conditions.
Furthermore, we find that axion inflation becomes warm over a large range of couplings, and explicitly map the parameter space for two axion inflation potentials.
We show that in large regions of parameter space, axion inflation models once assumed to be safely ``cold'' are in fact warm, and must be reevaluated in this context.

\end{abstract}

\maketitle

\section{Introduction}

Single-field slow roll inflation is a compelling solution to the difficulties of standard Big Bang cosmology~\cite{Linde:1981mu,PhysRevD.23.347} and provides a natural explanation for the primordial perturbations observed in Cosmic Microwave Background (CMB) anisotropies~\cite{Aghanim:2018eyx}. Realizing a viable model of single-field inflation however has challenges; the generation of radiative corrections to the potential by generic Planck-suppressed operators can spoil the flatness necessary to sustain a sufficient period of slow-roll (see e.g.~\cite{ArkaniHamed:2003mz} for a discussion). These corrections can be avoided if the inflaton is a pseudo-Nambu-Goldstone boson (pNGB) that enjoys a weakly-broken shift symmetry. As a result, such a pNGB (henceforth \textit{axion}) provides an elegant candidate for the inflaton. This was the original motivation for ``natural inflation''~\cite{PhysRevLett.65.3233} and the various axion inflation models that have since been proposed (e.g.~Refs.~\cite{Silverstein:2008sg,Berg:2009tg,Kaloper:2008fb,Dimopoulos:2005ac,Easther:2005zr,Kaplan:2003aj,Green:2009ds,Germani:2010hd,Anber:2009ua,Visinelli:2011jy,Adshead:2012kp,Maleknejad:2011jw,ArkaniHamed:2003wu,Caputo:2019joi} constitute a non-exhaustive list).

While the use of an axion as the inflaton avoids the issues of radiative corrections to the potential, it has other implications for the dynamics of inflation. Namely, the axion generically couples to other particles in the theory.  For example, such couplings are motivated by reheating~\cite{PhysRevD.60.023508,Finelli_2001,BLUMENHAGEN2014482,Adshead_2015,Halverson_2019}.  In particular, an axion, $\phi$, naturally couples to gauge fields with a dimension-5 operator $\propto\frac{\phi}{f}\mathrm{Tr}G\tilde{G}$ where $f$ is the scale of the symmetry breaking and $G$ is the gauge field strength.  We will show in this paper that such a coupling often generates a thermal bath during inflation, which can significantly alter the predictions of inflation.

As the axion rolls in its potential, it sources the production of gauge degrees of freedom. In some regions of parameter space, this production outpaces the dilution due to Hubble expansion and generates a thermal bath. Dissipative interactions with a thermal bath result in ``thermal friction'' that dramatically alters the inflationary dynamics. For non-Abelian gauge sectors, the friction is dominated by strong sphaleron transitions~\cite{Berghaus:2019whh,Laine:2021ego}, an understudied source of dissipation entirely distinct from those often considered~\cite{Anber:2006xt,Ferreira:2017lnd,PhysRevD.93.063510}.

Inflation in a coincidently-evolving thermal bath is known as ``warm inflation''~\cite{Berera:1995ie}. Warm inflation models often suffer from a similar problem to the aforementioned ``cold'' inflation models, though it is thermal corrections rather than Planck-suppressed radiative corrections that spoil the flatness of the potential~\cite{Yokoyama:1998ju}. Again, the axion avoids this, further motivating it as an inflaton in a warm inflationary setting~\cite{Ferreira:2017lnd,Visinelli:2011jy,Bastero-Gil:2016qru,Berghaus:2019whh,Kamali_2019}.

Most axion inflation models are however assumed to be in the ``cold regime'' where any backreaction from thermal effects is negligible. The purpose of this paper is to demonstrate that this is \textit{not} always a valid assumption.
As will be shown in the following sections, axion inflatons with gauge couplings must necessarily be exceptionally-weakly coupled to avoid being driven into a warm regime, whereupon observable predictions change dramatically. Furthermore, we show that in a large region of parameter space, this fate is independent of initial conditions. While in this paper we explore parameter space for two specific potentials of interest, the result we find is generic: axion inflation can often lead to the formation of a thermal bath that \textit{must} be accounted for.

This paper is organized as follows.  In Section~\ref{sec:review}, we review the dynamics of warm inflation in the context of an axion coupled to a non-Abelian gauge sector (with the possible presence of light fermions). In Section~\ref{sec:attractor}, we show that warm inflation is an attractor solution with a large basin of attraction.  In Section~\ref{sec:regimes}, we introduce the different possible regimes for warm and cold inflation.  In Section~\ref{sec:results}, we map out inflationary parameter space in terms of these regimes for two different cases:  one in which the axion couples to a pure $SU(3)$ Yang-Mills sector, and one in which the axion couples to the Standard Model $SU(3)_c$, so the effects of light fermions must be included.  We also analyze some of the qualitative features of the resulting maps of viable parameter space.  Finally, we conclude in Section~\ref{sec:conclusion}. Appendices~\ref{app:thermbath} and~\ref{app:vac} contain detailed computations of the bounds on initial conditions presented in Section~\ref{sec:attractor}.

\section{Inflation in a thermal bath}
\label{sec:review}

Axions couple to gauge sectors with a dimension-five non-renormalizable coupling of the form $\propto\frac{\phi}{f}\mathrm{Tr}G\tilde{G}$ where $f$ is the axion decay constant. 
While the potential is generally set by UV physics in axion inflation models, such a coupling to any gauge sector is allowed and generically expected to exist in the IR theory. In this case, the thermal bath that the motion of the inflaton sources can backreact and drive the system into the regime of warm inflation when naively, the system is thought to be cold.

Here, we consider two possible couplings: one to a generic $SU(3)$ gauge sector with no associated light fermions, the other to the strong gauge sector of the Standard Model. The Lagrangian includes the terms
\begin{multline}
\mathcal{L} \supset
\frac{1}{2}(\partial\phi)^2 - V(\phi)
-\frac{1}{2g^2}\text{Tr}G_{\mu\nu}G^{\mu\nu}
-\frac{\phi}{16\pi^2 f}\text{Tr}G_{\mu\nu}\tilde{G}^{\mu\nu}\\
-\left[\bar{\psi}(\slashed{D}-m_f)\psi\right]
\end{multline}
with $\phi$ the inflaton, $V(\phi)$ the inflationary potential set by UV physics, $g$ the gauge coupling of the non-Abelian gauge sector, $G^{\mu\nu}$ the associated field strength tensor, $\tilde{G}^{\mu\nu}=\frac{1}{2}\epsilon_{\mu\nu\rho\sigma}G^{\rho\sigma}$ its dual, and $f$ the decay constant which couples the inflaton to the gauge sector. We include the final term for models with a light fermion $\psi$ with mass $m_f$, such as the Standard Model, but we do not include this term when considering our pure gauge case.

The equations of motion for a homogeneous inflaton $\phi(t)$ in the presence of a thermal bath at temperature $T(t)$ are given by~\cite{Goswami:2019ehb}
\be
\ddot{\phi} + 3H\dot{\phi} + \Upsilon(T)\dot{\phi} + V'(\phi) = 0
\ee

\be
H^2 - \frac{1}{3M_{\text{pl}}^2}\left(\rho_R + \frac{\dot{\phi}^2}{2} + V(\phi) \right) = 0
\ee

\be
\label{eq:rhoR}
\dot{\rho}_R + 4 H \rho_R - \Upsilon(T)\dot{\phi}^2 = 0
\ee
where we have defined the energy density of the thermal bath of radiation as $\rho_R = \frac{\pi^2 g_*}{30}T^4$ with $g_*$ the number of relativistic degrees of freedom in the bath. $M_{\text{pl}} = \sqrt{\hbar c/8 \pi G} = 2.435\times10^{18}$ GeV/$c^2$ is the reduced Planck mass.

$\Upsilon(T)$ is the temperature-dependent dissipation term that accounts for the frictional effect of sphaleron transitions between gauge vacua. This has been computed in the generic case of a gauge sector (with trace normalization $T_R$ and dimension $d_R$ of the representation) and an associated light fermion to be~\cite{Berghaus:2020ekh}:
\be
\label{eq:ups}
\Upsilon(T) = \frac{\Gamma_{\text{sph}}}{2Tf^2}\left(\frac{\Gamma_{\text{ch}}}{\Gamma_{\text{ch}}+\frac{24 T_R^2}{d_R T^3}\Gamma_{\text{sph}}}\right)
\ee
where the sphaleron transition rate has been found to be
\be
\Gamma_{\text{sph}} \sim N_c^5 \alpha^5 T^4
\ee
with $N_c$ the number of colors in the sector and $\alpha = g^2/4\pi$ the coupling; the chirality-violation rate associated with the fermion is\footnote{For theories with multiple light fermions, it is the mass of the lightest fermion which will appear in Eq.~\ref{eq:chiral}.} 
\be
\label{eq:chiral}
\Gamma_{\text{ch}} \equiv \frac{\kappa N_c \alpha m_f^2}{T}
\ee
with $\kappa$ some $\mathcal{O}(1)$ coefficient which we set to 1 henceforth.

The effect of light fermions is to allow chirality-violating processes that diminish the friction associated with sphaleron transitions (see~\cite{Berghaus:2020ekh} for details). It is clear that in the limit $m_f\rightarrow \infty$, Eq.~\ref{eq:ups} reduces to the form previously derived for a pure-gauge sector~\cite{Berghaus:2019whh}. When $m_f \lesssim (N_c\alpha)^2 T$, the pure-gauge frictional term is suppressed by $\propto (N_c\alpha)^{-4}(m_f/T)^2.$

Note that the above expressions only hold if thermalization of the bath is efficient. This occurs when the thermalization rate, given by~\cite{Laine:2021ego}
\be
\label{eq:thermrate}
\Delta \approx 10 N_c^2 \alpha^2 T,
\ee 
is much greater than the rate of expansion $H$. In the relevant parameter space, this reduces to the condition $T/H \gtrsim 1$, which is equivalent to being in a warm inflation regime.

The slow roll conditions in warm inflation are modified from those of cold inflation. In cold inflation, one defines the \textit{slow roll} parameters $\epsilon_V$ and $\eta_V$ as
\be\label{eq:epsV}
\epsilon_V = \frac{1}{2}\left(\frac{V'}{V}\right)^2 M_{\text{pl}}^2
\ee
and
\be
\eta_V = \left(\frac{V''}{V}\right) M_{\text{pl}}^2.
\ee
In warm inflation, these equations are modified to 
\be
\epsilon_W = \frac{1}{(1+Q)}\frac{1}{2}\left(\frac{V'}{V}\right)^2 M_{\text{pl}}^2
\ee
and
\be
\eta_W = \frac{1}{(1+Q)}\left(\frac{V''}{V}\right) M_{\text{pl}}^2.
\ee
where $Q \equiv \Upsilon/3H$ is the ratio of the thermal friction to the Hubble friction. Clearly, these reduce to the cold parameters in the limit that $Q \ll 1$.  We refer to the regime in which $T > H$ but $Q\ll1$ as  \emph{weak} warm inflation, while we refer to the limit $Q\gg 1$ as \emph{strong} warm inflation.

The slow-roll conditions therefore become $\epsilon_W, \eta_W \ll 1$. Under these conditions, the equations of motion simplify to
\be\label{eq:eom1}
\dot{\phi} = \frac{-V'(\phi)}{3H + \Upsilon(T)}
\ee
\be\label{eq:eom2}
H^2 = \frac{\rho_R + V(\phi)}{3 M_{\text{pl}}^2}
\ee
\be\label{eq:eom3}
4H\rho_R = \Upsilon(T)\dot{\phi}^2
\ee
For any given potential $V(\phi)$ and axion decay constant $f$ (which appears in the expression for $\Upsilon$), these equations can be solved numerically to find the steady-state equilibrium configuration.

Inflation terminates when $\epsilon_W = 1$. Denoting the field value at which this occurs as $\phi_{\text{end}}$, one can then compute the number of e-folds of inflation between a given $\phi$ and $\phi_{\text{end}}$ as
\be
N = \frac{1}{M_{\text{pl}}^2}\int_{\phi_{\text{end}}}^{\phi} (1+Q(\tilde{\phi}))\left(\frac{V(\tilde{\phi})}{V'(\tilde{\phi})}\right)~d\tilde{\phi}.
\ee
By setting $N = N_*$ (taken to be 60 in this analysis\footnote{While the number of e-folds ultimately depends on the inflationary scale of a particular model, we choose to fix $N_*$ to a particular fiducial value for comparative purposes. We have verified that the maps of  parameter space we present in Sec.~\ref{sec:results} do not change appreciably for other choices of $N_*$.}), 
one can compute $\phi_*$, the field value at which one wishes to match theoretical predictions to cosmological observables. In this analysis, we will be interested in the equilibrium values of $H$, $T$, and $\Upsilon$ at this $\phi_*$.

\section{Warm Inflation as an Attractor}
\label{sec:attractor}

In this section, we will show that the equilibrium configuration derived in the previous section is an attractor solution with a large basin of attraction. This further underlines the key point of this paper: generic conditions in axion inflation often lead to thermal effects that cannot be ignored. We will explore two generic initial conditions for inflation and demonstrate that they both may lead to the warm inflation equilibrium. The first is a thermal bath at a temperature different than the equilibrium temperature,  the second is a vacuum absent of any thermalized degrees of freedom.

\subsection{Inflation begins in a thermal bath}
\label{sec:thermbath}

We will begin with the simple assumption that the inflationary period is preceded by some matter- or radiation-dominated period such that there is a well-defined initial temperature of the bath at the onset of inflation. Let $T_0$ denote the initial temperature of the bath and $T_{\text{eq}}$ be the steady-state equilibrium temperature of the system arising from Eqs.~\ref{eq:eom1}--\ref{eq:eom3}.

If $T_0 > T_{\text{eq}}$, the approach to equilibrium is simple to understand: the redshifting of the existing thermal bath is faster than the rate at which thermal friction dumps energy into the bath, thus Hubble expansion will cause the temperature of the bath to fall with $a^{-1}$, where $a$ is the scale factor.  In this case the system reaches the equilibrium temperature in $\sim \ln(\frac{T_0}{T_{\text{eq}}})$ e-folds of inflation. 

If $T_0 < T_{\text{eq}}$, the system will still reach equilibrium within a Hubble time, as first noted in Ref.~\cite{Berghaus:2019whh}.  The proof is as follows. We will assume that thermalization is efficient and the system remains at a well-defined temperature throughout its evolution. (In Appendix~\ref{app:thermbath}, we explicitly check these assumptions by ensuring that both $\Delta \gg \dot{T}/T$ and $\Delta \gg H$ throughout the evolution.) Then the evolution of $T$ is governed by Eq.~\ref{eq:rhoR}, along with the definition of $\rho_R$ and expression for $\Upsilon(T)$.  To encompass both the cases with and without light fermions, let us parameterize the dissipation as $\Upsilon(T)=CT^p$ (where $p=3$ when no light fermions are present and $p=1$ when there are).  As we will eventually see, the system will reach its equilibrium within a Hubble time, so the $4H\rho_R$ term in Eq.~\ref{eq:rhoR} can be neglected.  Substituting in the definition of $\rho_R$ and expression for $\Upsilon(T)$, we then find
\be\label{eq:Tevolve}
\dot\rho_R\approx\Upsilon(T)\dot\phi^2\implies T^{3-p}\dot T\approx\frac{15C}{2\pi^2g_*}\dot\phi^2.
\ee

Since $\dot\phi$ decreases as the temperature increases, it will be smallest when the system reaches its equilibrium.  We can therefore bound $\dot\phi>\dot\phi_\text{eq}$.  Then we can integrate the above to find
\be\label{eq:Teqbound}
\frac{T_\text{eq}^{4-p}-T_0^{4-p}}{4-p}>\frac{15C}{2\pi^2g_*}\dot\phi_\text{eq}^2t_\text{eq}
\ee
where $t_\text{eq}$ is the time it takes to reach its equilibrium.  Note that we can neglect the $T_0^{4-p}$ term here and this inequality will still be satisfied. We can derive an expression for $\dot\phi_\text{eq}$ by considering the equilibrium limit of Eq.~\ref{eq:rhoR}.  Then $\dot\rho_{R,\text{eq}}=0$, so we find
\be
\frac{2\pi^2g_*}{15}HT_\text{eq}^4=CT_\text{eq}^p\dot\phi_\text{eq}^2
\ee
Substituting this expression for $\dot\phi_\text{eq}$ into Eq.~\ref{eq:Teqbound} yields
\be
\frac{T_\text{eq}^{4-p}}{4-p}>HT_\text{eq}^{4-p}t_\text{eq}\implies t_\text{eq}<\frac1{(4-p)H}.
\ee
Therefore the system reaches equilibrium within a Hubble time so long as $p<4$, which is true for both the case with and without light fermions.  (This retroactively justifies neglecting the $4H\rho_R$ term when deriving Eq.~\ref{eq:Tevolve}.)

\subsection{Inflation begins in a vacuum}
\label{sec:vac}

Now let us consider a situation in which inflation begins in a vacuum.\footnote{This is in some sense a less generic case than those addressed in the previous subsection, as inflationary models are often preceded by an epoch in which radiation or matter dominate. It is in fact one of the original motivations of inflation that the exponential increase in the scale factor could inflate away the degrees of freedom of a previous epoch.} There is no well-defined $T_0$, hence the above argument does not immediately apply. However, if some other mechanism can generate gauge modes with energy density $\rho_R >\frac{\pi^2g_*}{30}(\frac{HT}\Delta)^4$, one would generically expect self-interactions of the non-Abelian gauge sector to thermalize them, leading to a bath with temperature $T$ and thermalization rate $\Delta$ greater than $H$.  Then the above proof shows that the system will run to the attractor solution. As we will see in the remainder of this section, the exponential production of gauge modes during inflation could do just that.

The exponential production of gauge modes during inflation arises due to a tachyonic instability in the gauge equations of motion~\cite{Campbell:1992hc, Anber:2009ua}. Neglecting the expansion of the universe, nonlinear terms arising from the non-Abelian nature of the $SU(3)$ gauge fields, and backreaction on the inflaton (assumptions that we will return to momentarily), the equations of motion for the gauge modes can be written as
\begin{equation}\label{eq:gaugeeom}
\ddot A_\pm-k\left(\frac{\alpha\dot\phi}{2\pi f}\mp k\right)A_\pm=0.
\end{equation} where $A_{\pm}$ are the helicity modes. It is immediately clear that for the $+$ polarization in the case where $\dot\phi>0$, modes with $k\leq k_\text{max}\equiv\frac{\alpha\dot\phi}{2\pi f}$ exhibit a tachyonic instability.

The gauge modes will grow under this mechanism until one of the above assumptions is violated.  In particular, if the linearity assumption is the first to be violated, nonlinear dynamics will take over.  While we cannot say for sure what happens to the gauge modes at that point, on the basis of entropic arguments we find it likely that they thermalize.  (Simulations like those done in the $U(1)$ context~\cite{PhysRevD.93.063510} or on the lattice during preheating~\cite{Adshead_2017} would be useful to confirm this intuition.)  We thus wish to determine under what conditions the linearity assumption is the first to be violated, and that the mechanism has produced enough modes by that time to thermalize into a bath with $\Delta > H$.  We must therefore ensure three conditions are satisfied:
\begin{enumerate}
    \item \textit{Do nonlinearities become relevant within a Hubble time?} This ensures that the mechanism produces enough gauge modes for thermalization to begin before they are diluted by Hubble expansion.
    \item \textit{Do nonlinearities become relevant before the production of gauge modes backreacts on the inflaton's motion?} We must ensure that the exponential production does not drain the inflaton of its kinetic energy before it has produced enough modes to thermalize.
    \item \textit{Is there enough energy density in the gauge modes when nonlinearities become relevant to source a bath with $\Delta > H$?} This ensures that this mechanism can source a bath with a temperature large enough for the arguments of the previous section to apply.
\end{enumerate}
The analytic derivation of these conditions is tedious, and can be found in Appendix~\ref{app:vac}. In this section, we will simply state the results, though we refer the interested reader to the Appendix for a full treatment.

\textbf{Condition 1:} We require that nonlinearities become large well before Hubble expansion can dilute the gauge modes. This will be satisfied when
\begin{equation}
\label{eq:cond1}
\frac{f}{M_\text{pl}} \ll \mathcal{O}(10^{-1})\alpha\sqrt{\epsilon_V}.
\end{equation}
We have computed this bound, as well as the following bounds, by setting the inflaton's velocity to the terminal velocity set by Hubble friction, i.e. $\dot{\phi} = V'(\phi)/3H$, however the bounds are also applicable in the case of lower velocities (see footnote~\ref{ftnt:velocity} of Appendix~\ref{app:vac}). Eq.~\ref{eq:cond1} can also be written as $k_\text{max} \gg \mathcal{O}(1)H$, from which we see that so long as the fastest growing modes $k\sim k_\text{max}$ are sub-horizon, these modes will satisfy Condition 1.

\textbf{Condition 2:} This condition demands that the energy drained from the inflaton by the exponential production of gauge modes does not backreact to slow the roll of the inflaton. Backreaction is avoided when
\begin{equation}
\frac{f}{M_\text{pl}} \gg \mathcal{O}(10^{-2})\sqrt[4]{\alpha^3\epsilon_V}\sqrt{\frac H{M_\text{pl}}}.
\end{equation}
It should be stressed that the $H$ and $\epsilon_V$ that appear in this expression correspond to the \textit{initial values} of these parameters. This makes them somewhat arbitrary, as one has the freedom to choose initial conditions. However, in order to provide some context for the regions of parameters space this bound covers, one can rewrite the expression as
\begin{multline}
f \gg 5\times10^{10}~\text{GeV}\\
\times\left(\frac{\alpha}{0.1}\right)^{3/4}
\left(\frac{\epsilon_V}{10^{-3}}\right)^{1/4}\left(\frac{H}{10^{-10}\,M_{\mathrm{pl}}}\right)^{1/2}
\end{multline}
(see Appendix~\ref{app:vac}).  It is clear that this can be satisfied for a wide variety of initial conditions in interesting regions of parameter space.

\textbf{Condition 3:} Finally, here we ensure that enough energy density is present in the gauge modes before nonlinearities take effect in order to form a bath with $\Delta > H$.  This occurs so long as
\begin{equation}
\frac{f}{M_{\text{pl}}} \ll \mathcal{O}(1)N_c^2\alpha^{11/4}g_*^{-1/4}\epsilon_V^{1/2}
\end{equation}
with $g_*$ the number of degrees of freedom in the bath.
Note that for $N_c^2\sqrt[4]{\alpha^7/g_*}\lesssim10^{-1}$, this is more constraining than Condition 1.

So, in summary, we find that provided the produced gauge modes are well sub-horizon (Conditions 1 and 3) and that $f$ is sufficiently large to avoid backreaction (Condition 2), the rolling inflaton will source a large energy density of gauge modes, even if the system begins from a vacuum.  This density will be sufficiently large that self-interactions will take effect, likely leading to its thermalization (although determining the exact process of thermalization is beyond the scope of this work).  If such thermalization occurs, the argument of the previous section would then apply to the bath, meaning the system will tend towards the warm inflation equilibrium. This suggests that there is a large region of parameter space in which warm inflation is an attractor solution, even if its initial condition is a vacuum.

\vspace{0.5cm}

The results of this entire section, in both the case of vacuum and initial thermal bath, underscore a major point of this paper: the fact that warm inflation is the attractor solution implies that if a warm inflationary equilibrium exists for a particular choice of parameters, the system will \textit{necessarily} be driven to it; for a large set of initial conditions, it cannot be evaded.

\section{Inflationary regimes}
\label{sec:regimes}

We are interested in what regions of parameter space ``cold'' inflationary models are in fact driven to warm inflation. As such, we chose to examine fiducial potentials that are popular in the axion inflation literature and calculate the consequences of an additional gauge coupling. The two potentials studied are that of axion monodromy~\cite{Silverstein:2008sg, McAllister:2008hb, Flauger:2009ab}
\be
V_{\text{mon}}(\phi) = \mu^3 \left(\sqrt{\phi^2 + \phi_c^2} - \phi_c \right)
\ee
and the Starobinsky model~\cite{STAROBINSKY198099} (which is a limit of the $\alpha$-attractors~\cite{Kallosh:2013yoa, PhysRevD.88.085038})
\be
V_{\text{Star}}(\phi) = V_0 \left(1 - \exp\left(-\frac{|\phi|}{v}\right)\right)^2.
\ee
Each is a two-parameter family of models governed by a normalization and scale: $\mu$ and $\phi_c$ respectively in monodromy, and $V_0$ and $v$ in the Starobinsky model. Best fits to the current measurements of the CMB parameters $A_s$, $n_s$, and $r$ by the Planck satellite~\cite{Aghanim:2018eyx} lead to parameter values of $\mu = 6\times10^{-4} M_{\text{pl}}$, $\phi_c = M_{\text{pl}}/10$ and $V_0 = 6.2\times10^{-10} M_{\text{pl}}^4$, $v = 10M_{\text{pl}}/3$ for the two models in the usual cold inflationary paradigm~\cite{Adshead:2019lbr}.

We note that it is however not possible to reproduce the measured CMB parameters in the strong warm regime with these potentials.\footnote{In the weak regime, the formulae for the spectral parameters depend on the phase-space distribution of inflaton particles, which we do not compute and thus refrain from commenting on. For a detailed discussion, see Ref.~\cite{Bartrum_2014}.} This is because correctly reproducing the observed red-tilt of the CMB in strong warm inflation requires $\eta_V\geq\epsilon_V$~\cite{Berghaus:2019whh} (in contrast to the cold inflation scenario where potentials with $\epsilon_V\geq\eta_V$ are sought to reproduce the observed spectral tilt). As a result, this condition is not met in generic ``cold'' inflation potentials such as these.  For this reason, we restrict our analysis to classifying inflationary regimes rather than computing spectral parameters.  While we limit our analysis to these two potentials, the results we display below are fairly generic for slow-roll inflationary potentials, as will be discussed further below.

For these potentials, we vary the normalizations and axion decay constants, under both the assumption of a pure $SU(3)$ gauge group with no light fermions and with Standard Model (SM) fermions and a coupling to the SM strong sector, and find the associated equilibrium solution. Depending on the resulting values for $H$, $T$, and $\Upsilon$ at $\phi_*$, the solution is sorted into one of the following categories:
\begin{enumerate}
    \item \textbf{Strong warm inflation:} Here, $Q>1$ (equivalently $\Upsilon > 3H$), hence the friction is dominated by thermal effects and the dynamics are significantly changed from the naive cold prediction. Additionally, $H, T < f$ to ensure that the behavior of the axion is well-defined.
    \item \textbf{Weak warm inflation:} In this case, $Q < 1$ (equivalently $3H > \Upsilon$), hence we are not in the strong regime, however $H < T$. While the dynamics in this case are not influenced by thermal effects, the CMB power spectrum is now set by thermal fluctuations~\cite{BasteroGil:2011xd,2009JCAP...07..013G}. Again, $H, T < f$ for the reason above.
    \item \textbf{Cold inflation:} In this case, $T < H$, hence both the dynamics and resulting CMB spectrum are set purely by the Hubble parameter and the predictions of the inflation model are unchanged. Again, $H, T < f$.
    \item \textbf{Thermally broken ($H < f < T$):} We denote this as ``thermally broken'' in that the derived equilibrium temperature is above the axion decay constant, hence the above assumptions made on the dynamics of the axion are no longer valid. Naively, if operating purely under the assumption of cold inflation, this may seem like viable parameter space, however the generation of a thermal bath leads to a backreaction that renders the effective theory invalid.
    \item \textbf{Broken ($f < H$):} In this regime, the effective theory also breaks down, since the Hubble scale is above the axion decay constant. Note that this breakdown is not related to the temperature, hence the non-viability of these regions of parameter space could have been predicted prior to this study.
\end{enumerate}

The most relevant of these categories for us are 1, 2, and 4, namely strong warm inflation, weak warm inflation, and thermally broken $(H < f < T)$, because for all three of these regimes, the naive assumption that a particular model is operating in a cold inflationary regime is violated, and thermal effects \textit{must} be taken into account.

Note that there is an additional criterion that can influence the validity of the axion EFT, namely whether $\dot\phi$ is less than $f^2$. One might expect that $\dot\phi > f^2$ indicates a region in which non-perturbativity renders the EFT picture invalid. A hopeful model-builder might try to construct a UV theory which escapes this condition, and as a result, we choose not to include it as a separate region. However, as we will show in the following section, any such model-building buys one very little, as the vast majority of the parameter space in which $\dot\phi > f^2$ is already broken regardless.

\section{Results}
\label{sec:results}

We study both couplings to a generic pure $SU(3)$ gauge sector with no light fermions (that may, for example, constitute a portion of a dark sector) and couplings to the strong $SU(3)_c$ of the Standard Model, taking into consideration the effects of light fermions in the SM.

As is shown below, the monodromy and Starobinsky models display similar behavior for both of these cases. This is not surprising, as the relevant dynamics are in general occurring on scales much smaller than those on which the potential is changing, hence the results are fairly insensitive to the overall shape of the potential so long as the slow-roll conditions are satisfied. This indicates that this rough map of parameter space is generalizable to other models that employ a slow-rolling axion as the inflaton.

\subsection{Pure $SU(3)$}

For the pure gauge sector, we adopt QCD-like parameters, setting $\alpha = 0.1$ and $N_c = 3$. We assume that the new sector is sequestered from SM degrees of freedom and adopt $g_* = 2N_c^2 - 1 = 17$, which accounts for the two polarizations for each of the eight gauge bosons and a single degree of freedom from the inflaton. The theory is taken to be unconfined at all relevant scales. 
Finally, we choose to take $N_* = 60$ in the below plots, but the results are robust against changes to this parameter.

\begin{figure*}
  \centering
  \includegraphics[width=0.98\textwidth]{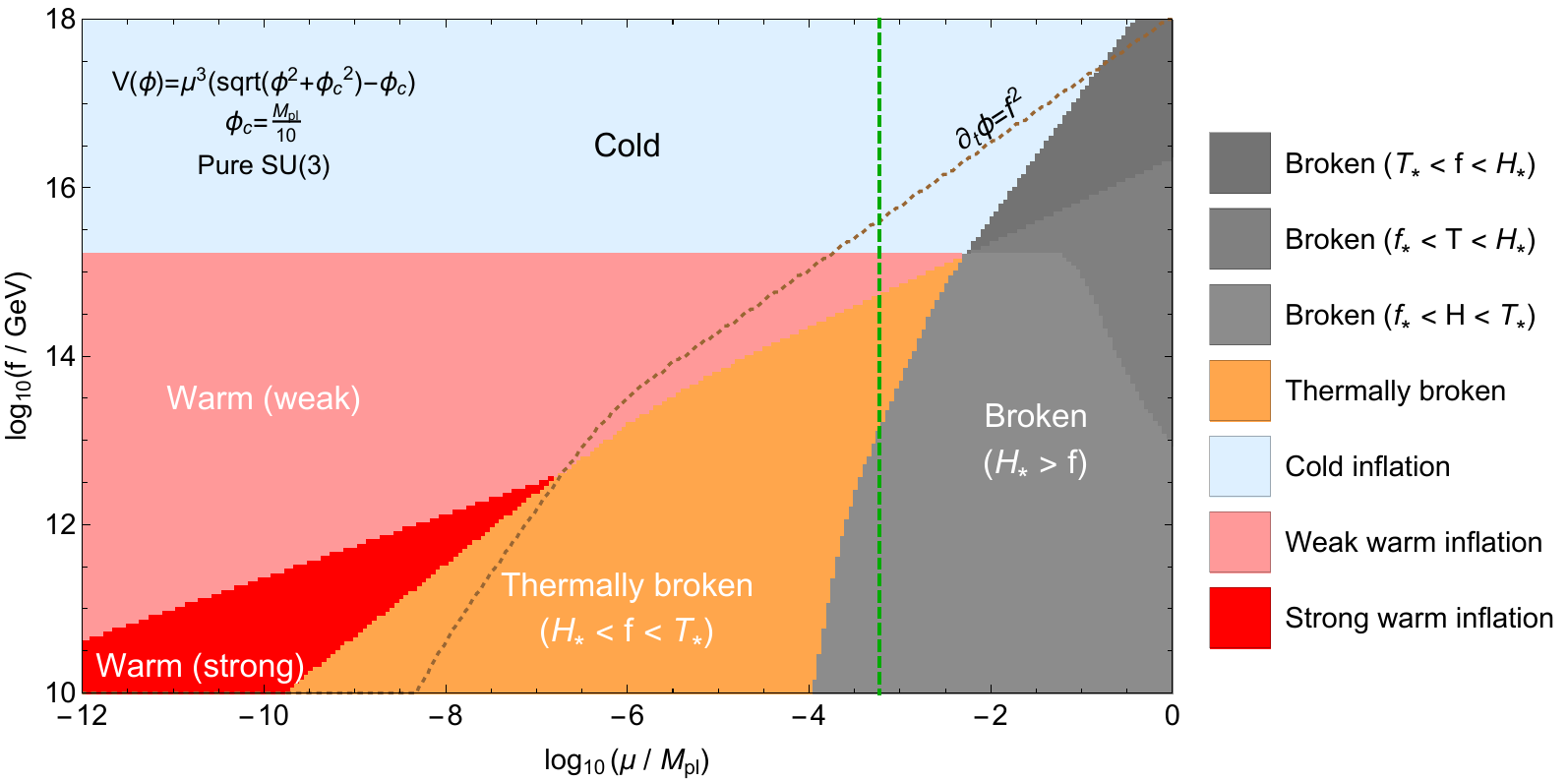}
  \caption{Map of parameter space for an axion monodromy potential with coupling to pure $SU(3)$ gauge sector (no light fermions). Light blue regions indicate parameter space in which inflation remains cold, while red regions indicate that inflation enters a warm regime. The orange region corresponds to the regime in which the temperature of the bath is driven above $f$, hence the EFT breaks down and cold inflation is certainly not a good assumption. Below the dotted brown line, $\partial_t\phi > f^2$, hence the validity of the EFT could be questioned. The dashed green line denotes the required normalization to reproduce the observed $A_s$ of the CMB power spectrum. It is clear that for $f \lesssim 10^{15}$ GeV, the assumptions of cold inflation are violated for this choice of normalization. Additionally, we note the appearance of large warm inflationary regimes towards small values of the normalization.
     \label{Fig:monoNoFerm}}
\end{figure*}

\begin{figure*}
  \centering
  \includegraphics[width=0.98\textwidth]{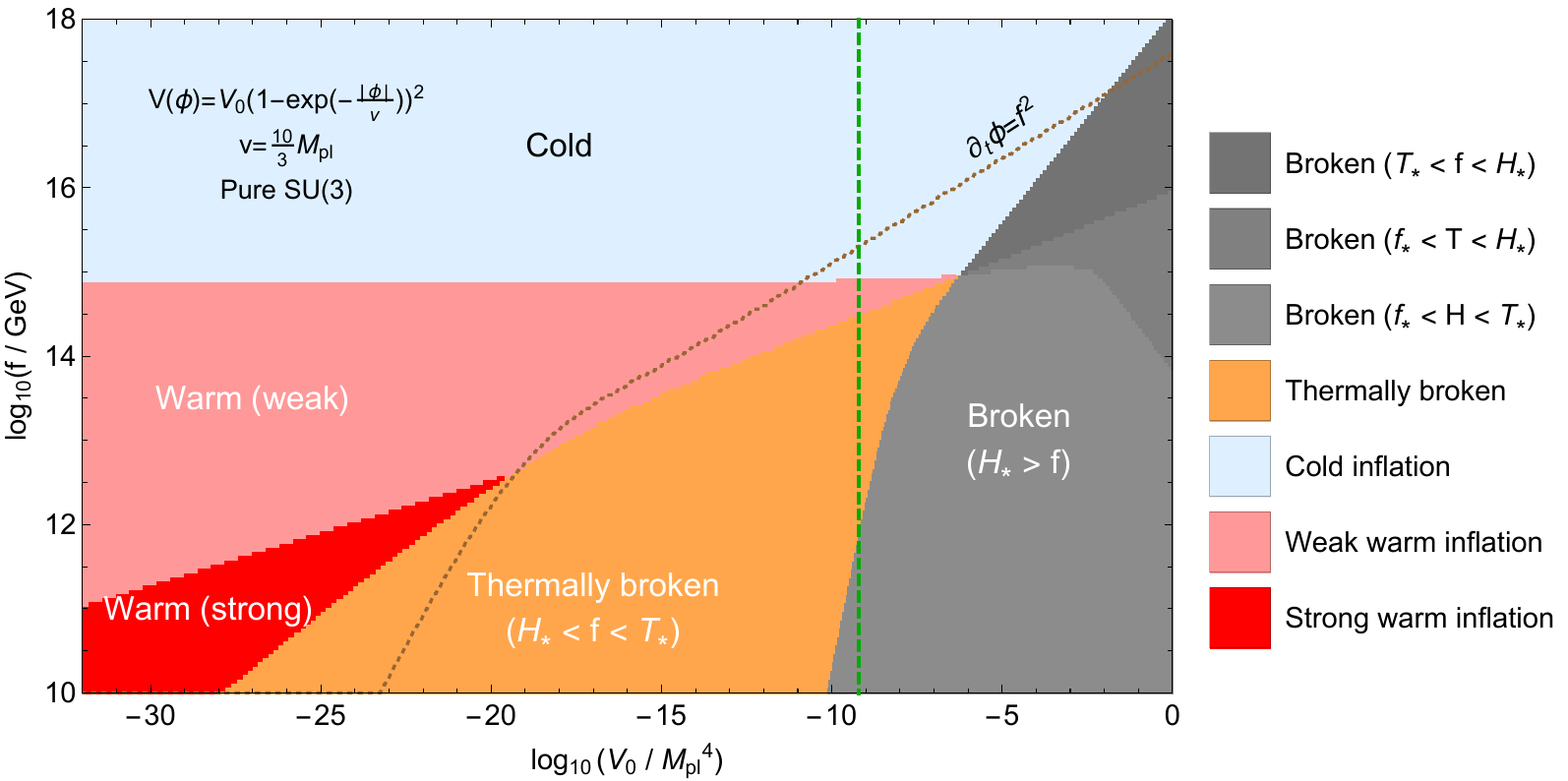}
  \caption{Same as Fig.~\ref{Fig:monoNoFerm} for the Starobinsky potential. As expected, this exhibits similar behavior to that of the monodromy potential.
     \label{Fig:starNoFerm}}
\end{figure*}

In Figures \ref{Fig:monoNoFerm} and \ref{Fig:starNoFerm}, we show the parameter space spanned by the axion decay constant and normalization of the potential. The dashed green line indicates the required normalization to match $A_s$ under the assumption of cold inflation, hence these cold inflation models are only consistent and viable in regions where this dashed green line lies within the light blue cold inflation regime. It is clear that in both cases, this limits the allowed axion couplings to $10^{-3} M_{\text{pl}} < f < M_{\text{pl}}$ in order for these models to be safe from thermal effects, whereas prior to this work one may have hoped to construct cold models with decay constants as low as $f\sim10^{-4} M_\text{pl}$.\footnote{Note that the point at which $f$ becomes smaller than $H$ is a bit misleading in Figs.~\ref{Fig:monoNoFerm} and~\ref{Fig:starNoFerm}. Since Hubble is evaluated at the $\phi_*$ predicted under warm conditions, the Hubble that \textit{would} have been predicted under the assumption of cold inflation is marginally larger.}

\subsection{The Standard Model}

\begin{figure*}
  \centering
  \includegraphics[width=0.98\textwidth]{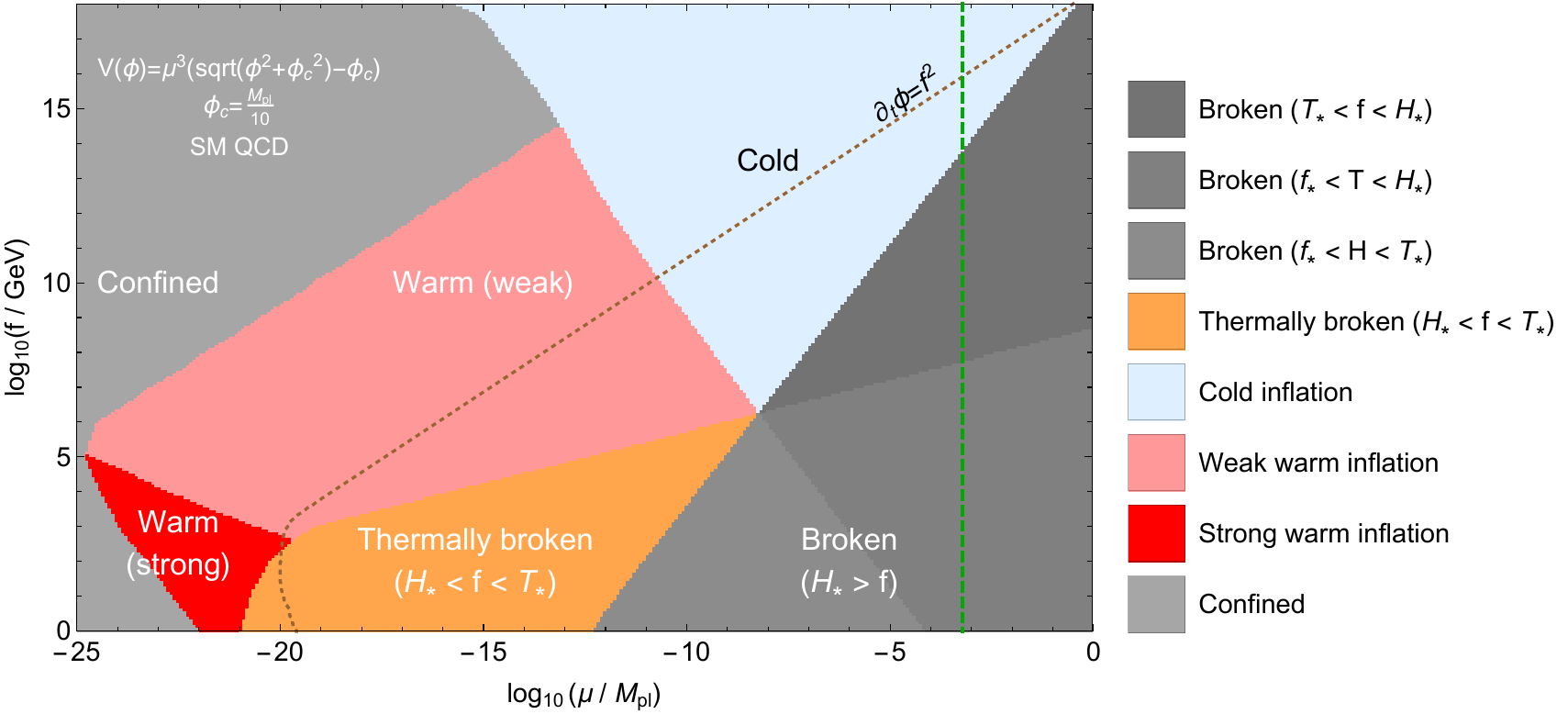}
  \caption{Map of parameter space for an axion monodromy potential coupled to the QCD sector of the Standard Model. The dashed green line denotes the required normalization to reproduce the observed $A_s$ of the CMB power spectrum. The inclusion of the effects of light fermions has dramatically changed the locations of the various regimes, weakening the frictional mechanism and pulling the $f=H=T$ point towards lower values of the coupling and normalization. It is interesting to note that the regions of viable warm inflation towards lower couplings persist even with this effect.
     \label{Fig:monoFerm}}
\end{figure*}

\begin{figure*}
  \centering
  \includegraphics[width=0.98\textwidth]{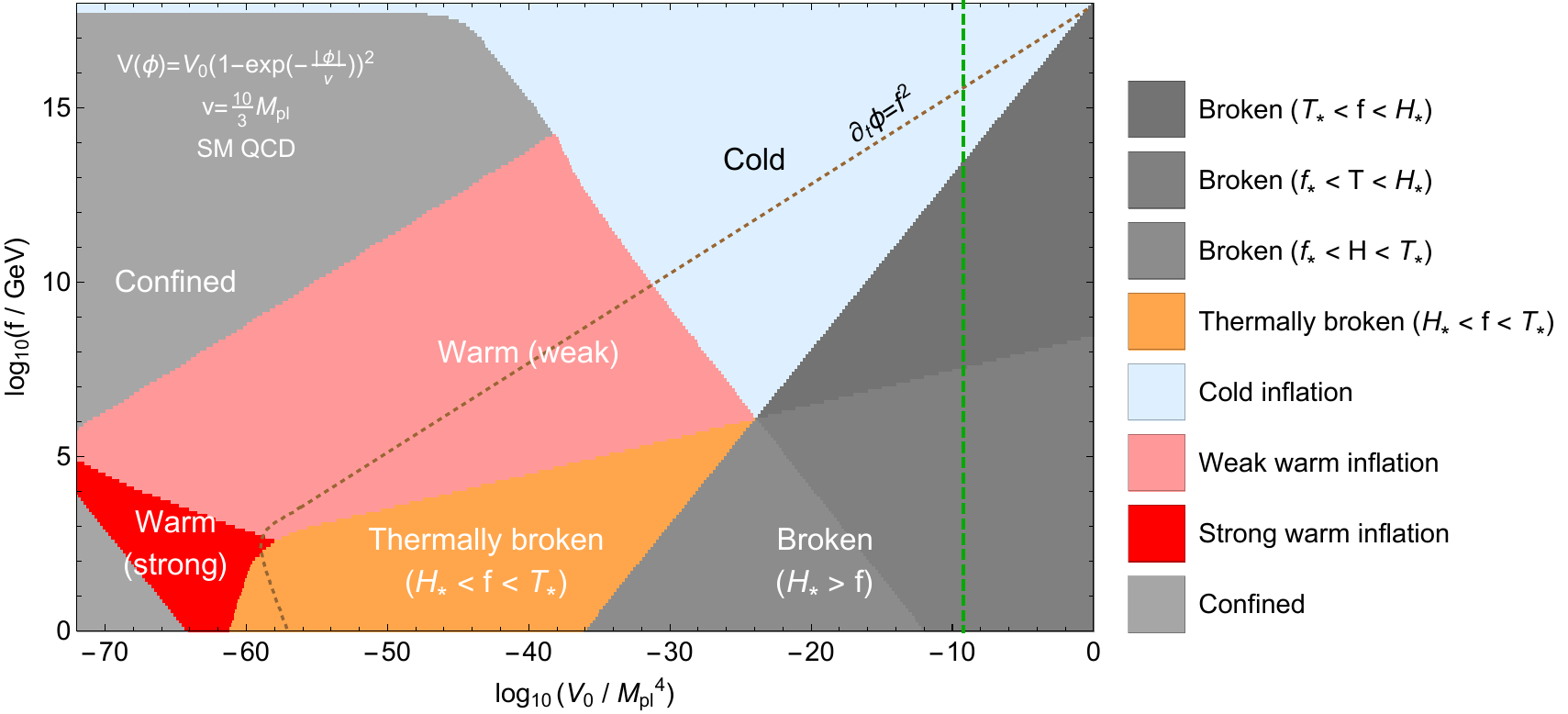}
  \caption{Same as Fig.~\ref{Fig:monoFerm} for the Starobinsky potential. As expected, this exhibits similar behavior to that of the monodromy potential.
     \label{Fig:starFerm}}
\end{figure*}

Here, we adopt the same parameters as in the previous section, however we change $g_* \rightarrow g_*(T)$ to account for all relativistic species in the Standard Model below temperature $T$~\cite{Husdal:2016haj} and additionally compute the running of $\alpha_{\text{QCD}}$ with temperature. We employ the fit in~\cite{Deur:2016tte} that sets $n_f = 5$ and fixes $\alpha_{\text{QCD}}(m_Z)$ at the mass of the $Z$ boson. This introduces a new effect, which is that for sufficiently low temperatures ($T\sim0.1$ GeV), the theory confines. We have demarcated these regions in light gray in Figs.~\ref{Fig:monoFerm} and \ref{Fig:starFerm} and labeled them accordingly. Additionally, we take $m_f \approx 1$ MeV in Eq.~\ref{eq:chiral}, along with $T_R=1/2$ and $d_R=3$ in Eq.~\ref{eq:ups}, to account for the light fermionic degrees of freedom in the Standard Model. 

As above, we show the parameter space spanned by the axion decay constant and normalization of the potential. Once again, the dashed green line indicates the required normalization to match $A_s$ under the assumption of cold inflation, hence these cold inflation models are only consistent and viable in regions where this dashed green line lies within the light blue cold inflation regime.

In comparison to the previous subsection, it is clear that the light fermions play a significant role in limiting thermal effects by reducing the friction and the associated energy density of the thermal bath for a given Hubble. The crossing point at which $f = T = H$ has been shifted to significantly lower values of $f$ and normalization, hence lower $H$. As a result, for the correct normalization to reproduce $A_s$, no choice of $f$ forces the model into a region in which thermal effects dominate.

It should be noted that constraints on axion-gluon couplings~\cite{Alonso-Alvarez:2018irt,Graham:2013gfa} may be relevant at low normalizations and strong couplings (lower left of Figures~\ref{Fig:monoFerm} and \ref{Fig:starFerm}), however these are model-dependent. As placing limits is not the main point of this paper, we choose not to display these bounds.

\subsection{Regime Scalings}

Much of the rough structure of Figs.~\ref{Fig:monoNoFerm}--\ref{Fig:starFerm} is generic for all potentials and can be derived analytically.  This is because in the weak limit $Q\ll1$, both $\phi_\text{end}$ and $\phi_*$ are independent of $f$ and the normalization of the potential (since $\epsilon_W$ is independent of these).  Therefore $H(\phi_*)$ is proportional to the square root of the normalization of the potential, and thus is a good proxy for the horizontal axes of Figs.~\ref{Fig:monoNoFerm}--\ref{Fig:starFerm}.  Using Eq.~\ref{eq:ups} and Eqs.~\ref{eq:eom1}--\ref{eq:eom3}, it is then straightforward to derive how $T(\phi_*)$ and $Q(\phi_*)$ scale with $H(\phi_*)$ and $f$.  These can then be used to determine the slopes of the boundaries between these regions, which are defined by equating different sets of these parameters.

For example, the boundaries between the cold and weak warm regimes in Figs.~\ref{Fig:monoNoFerm}--\ref{Fig:starNoFerm} appear as horizontal lines.  This behavior can be derived analytically by plugging the appropriate limits into Eq.~\ref{eq:eom3}.  Namely since there are no light fermions, we may take the $m\rightarrow\infty$ limit of Eq.~\ref{eq:ups} and find $\Upsilon\sim T^3/f^2$.  Additionally since we are interested in the weak limit, we may ignore the $\Upsilon$ in the denominator of Eq.~\ref{eq:eom1} to find $\dot\phi\sim V'/H\sim H$.  Thus in these limits Eq.~\ref{eq:eom3} yields $T\sim H/f^2$.  The boundary between the cold and weak warm regimes is defined by $T=H$, and thus will occur at a constant value of $f$ in these limits.

We can likewise understand why this behavior changes in Figs.~\ref{Fig:monoFerm}--\ref{Fig:starFerm}.  The bend in the boundary between the cold and weak warm regimes is due to the change in the relevant limit of Eq.~\ref{eq:ups}.  Namely on the left-hand side of the plot, the $m_f\rightarrow\infty$ limit applies, and so the line is horizontal as in the previous case.  However on the right-hand side, $m_f\ll\alpha^2T$ and so the appropriate scaling for the friction is $\Upsilon\sim T/f^2$.  By a similar argument to above, this implies that the boundary at $T=H$ should scale as $f\sim H^{-1}$, which is reflected in the Standard Model plots.  Note that the bend in the boundary occurs roughly when $H\approx\sqrt{\frac{d_R\kappa}{24}}\frac{m_f}{N_c^2\alpha^2T_R}$, since this is the temperature at which Eq.~\ref{eq:ups} transitions between these two limits.

Similar arguments to the above may be made to understand many of these boundaries.  We remark on a few interesting ones and give qualitative explanations of them.  Firstly, note that the boundary of the confined region changes direction around the transition from the weak warm regime to the strong warm regime.  Since the boundary of the confined region is approximately defined by a fixed temperature, this means that in the weak limit, decreasing $f$ (going to stronger couplings) increases the temperature, while in the strong limit, it decreases the temperature.  This is simply because in the weak limit, the thermal friction does not influence the dynamics of inflation.  Thus decreasing $f$ does not slow the inflaton down and can only cause more energy to be dumped into the bath.  In the strong limit, however the thermal friction dominates the dynamics, so decreasing $f$ will slow the inflaton down, leading to less energy dumped into the bath and a lower temperature.

Next note that the boundaries between the weak warm and strong warm regimes have opposite slopes between Figs.~\ref{Fig:monoNoFerm}--\ref{Fig:starNoFerm} and Figs.~\ref{Fig:monoFerm}--\ref{Fig:starFerm}.  This is due to the different scalings of Eq.~\ref{eq:ups} in the cases with and without light fermions.  (The boundaries in Fig.~\ref{Fig:monoFerm}--\ref{Fig:starFerm} lie in the $m_f\ll\alpha^2T$ limit, so the case with light fermions is appropriate.)  Using Eqs.~\ref{eq:eom1} and \ref{eq:eom3}, it can be shown that when $\Upsilon=H$, then $H\sim T^2$.  Thus in the case without light fermions $\Upsilon\sim T^3/f^2$ scales faster than $H$ with temperature, while in the case with light fermions $\Upsilon\sim T/f^2$ scales slower.  This leads to opposite slopes in the two cases.

Finally, we point out that this technique of using $H(\phi_*)$ as a proxy for the horizontal axis breaks down in the strong limit, as $\phi_*$ can gain dependence on $f$.  For instance, if these figures were actually plotted with $H(\phi_*)$ on the horizontal axis, then the boundary between the broken and thermally broken regimes, which is defined by $f=H$, would appear as a straight line.  In Figs.~\ref{Fig:monoNoFerm}--\ref{Fig:starNoFerm}, this is clearly not true as majority of the boundary lies well within the strong regime.

\section{Conclusion}
\label{sec:conclusion}

The axion remains a well-motivated candidate for the inflaton as its potential is protected from radiative and thermal corrections that would spoil the slow-roll dynamics. However, we have shown that generic couplings to non-Abelian gauge sectors, motivated for example by reheating, can result in the generation of a thermal bath that significantly alters the predictions of inflation.
This bath is an attractor solution in significant parts of axion inflation parameter space.  In those regions, axion inflation will generate the equilibrium bath for a wide range of initial temperatures, and it seems likely that the bath would even be generated from an initial vacuum.  In other words, for those parts of inflation model parameter space, the presence of a thermal bath in axion inflation is \textit{not optional} and thermal effects cannot be neglected when assessing the viability of an axion inflation model. We find the general result that the coupling to a generic $SU(3)$ gauge group must be very weak ($f \gtrsim 10^{15}$ GeV) in order to remain safely in a cold inflationary regime, though this limit is modified in the presence of light fermions such as those that appear in the Standard Model.  The parts of parameter space that become warm for two representative axion inflation potentials are shown in Figures \ref{Fig:monoNoFerm} and \ref{Fig:starNoFerm} for the pure gauge case and Figures \ref{Fig:monoFerm} and \ref{Fig:starFerm} for the case with light fermions. Our results serve as a warning to inflationary model-builders: ``cold'' axion inflation is often warmer than one might expect.

\begin{acknowledgments}
The authors would like to thank a large number of people for useful discussions during the completion of this work, including David Kaplan, Kim Berghaus, Guy Moore, Dan Green, Victor Gorbenko, Eva Silverstein, Andrei Linde, Mehrdad Mirbabayi, and Azadeh Maleknejad, among others. In particular, the authors wish to thank Junwu Huang for useful comments regarding the first draft of this paper. The authors further wish to acknowledge the support provided in part by the Simons Investigator Award 824870, DOE Grant DE-SC0012012, NSF Grant PHY-2014215, DOE HEP QuantISED award \#100495, and the Gordon and Betty Moore Foundation Grant GBMF7946.  This work was also supported by the U.S. Department of Energy, Office of Science, National Quantum Information Science Research Centers, Superconducting Quantum Materials and Systems Center (SQMS) under the contract No. DE-AC02-07CH11359. SK is further supported by NSF Grant DGE-1656518. Some of the computing for this project was performed on the Sherlock cluster. We would like to thank Stanford University and the Stanford Research Computing Center for providing computational resources and support that contributed to these research results.
\end{acknowledgments}

\begin{appendix}

\section{Thermalization of an initial thermal bath}
\label{app:thermbath}

In Section~\ref{sec:thermbath}, we assumed that thermalization was efficient and the temperature was well-defined throughout its evolution.  This will be true so long as the thermalization rate $\Delta \sim 10 N_c^2 \alpha^2 T$ (Eq.~\ref{eq:thermrate}) satisfies $\Delta\gg\dot T/T$ and $\Delta\gg H$, which amount to conditions on thermalizing faster than the bath's temperature is evolving and faster than the bath is diluting due to the expansion of the Universe.  Since $T$ increases and $\dot T$ decreases as the evolution continues, it is sufficient to check both these conditions at $T_0$. The situation is quantitatively different in the presence/absence of a light fermion, hence we explore both, however the qualitative picture remains the same.

\subsection{No light fermions}

Let us first consider the case of a pure $SU(N)$ gauge sector with no associated light fermions, so that we may set $p=3$ and $C=N_c^5\alpha^5/2f^2$.  Eq.~\ref{eq:Tevolve} then implies
\be
\dot T_0\approx\frac{15N_c^5\alpha^5\dot\phi_0^2}{4\pi^2g_*f^2}
\ee
We may assume that $\dot\phi_0$ is no larger than its cold inflation value $-V'/3H$, as if it were it would be slowed down within a Hubble time (and we may just consider $T_0$ to be the temperature when $\dot\phi$ reaches its cold inflation value).  Then utilizing Eq.~\ref{eq:epsV} and Eq.~\ref{eq:eom2} (neglecting the $\rho_R$ contribution), we find
\be
\dot T_0\approx\frac{5N_c^5\alpha^5V'^2}{12\pi^2g_*H^2f^2}\approx\frac{15N_c^5\alpha^5\epsilon_VH^2M_\text{pl}^2}{2\pi^2g_*f^2}.
\ee
The condition $\Delta\gg\dot T/T$ then becomes
\be\label{eq:T0thresh}
10N_c^2\alpha^2T_0\gg\frac{\dot T_0}{T_0}\implies T_0\gg\sqrt{\frac{3N_c^3\alpha^3\epsilon_V}{4\pi^2g_*}}\frac{HM_\text{pl}}f.
\ee
Though initial conditions can be selected arbitrarily, we can evaluate Eq.~\ref{eq:T0thresh} at the set of fiducial values that we use in Section~\ref{sec:results} (namely $\alpha = 0.1$, $N_c = 3$, and $g_* = 2N_c^2 - 1 = 17$) to provide a numerical frame of reference:
\be\label{eq:T0vals}
 \frac{T_0}{H}\gg 3\times10^{-4}\left(\frac{N_c\alpha}{3\times0.1}\right)^{3/2}\left(\frac{\epsilon_V}{10^{-3}}\right)^{1/2}\left(\frac{f}{M_\text{pl}}\right)^{-1},
\ee
indicating a large region of parameter space in which this is satisfied.

\subsection{Light fermions}

For the case with light fermions, we instead take $p=1$ and $C=\frac{\kappa N_c\alpha d_R m_f^2}{48T_R^2f^2}$.  We then find
\begin{equation}
    T_0^2\dot T_0\approx\frac{5\kappa N_c\alpha d_R\epsilon_Vm_f^2H^2M_\text{pl}^2}{16\pi^2g_*T_R^2f^2},
\end{equation}
which yields the condition
\begin{equation}\label{eq:T0thresh_fermion}
    T_0\gg\sqrt[4]{\frac{\kappa d_R\epsilon_V}{32\pi^2g_*N_c\alpha T_R^2}}\sqrt{\frac{m_fHM_\text{pl}}f}.
\end{equation}
As above, we note that initial conditions are arbitrary, however we can evaluate this at a set of fiducial parameters (now additionally taking $g_* = 100$, $T_R = 1/2$, $d_R = 3$, and $m_f = 1$ MeV) to provide a numerical frame of reference. This yields
\begin{multline}\label{eq:T0vas_fermion}
    \frac{T_0}{H} \gg7\times10^{-13}\\
    \times\left(\frac{N_c\alpha}{3\times0.1}\right)^{-1/4}\left(\frac{\epsilon_V}{10^{-3}}\right)^{1/4}\left(\frac{H}{f}\right)^{-1/2}\left(\frac{f}{M_\text{pl}}\right)^{-1}
\end{multline}
which once again demonstrates the large basin of attraction.

In summary, in order for our proof that the system reaches the warm inflation equilibrium solution within a Hubble time to be valid, the bath must thermalize more rapidly than it dilutes due to Hubble expansion and thermalize more rapidly than the temperature changes. The first condition ($\Delta \gg H$) implies that $T_0$ must exceed $(10 N_c\alpha)^{-2} H$, and the second ($\Delta \gg \dot{T}/T$) implies that $T_0$ must exceed the threshold in Eq.~\ref{eq:T0thresh}/\ref{eq:T0thresh_fermion} in the absence/presence of fermions.  Note that this latter threshold is not a physical bound on reaching warm inflation, but rather a limit on the validity of our assumptions: when this condition is violated, the dissipation is \textit{so efficient} that the bath does not have time to thermalize all of the particles being produced. However, since a thermal distribution is ultimately the distribution of maximal entropy, it is very likely that the system eventually reaches a thermal equilibrium at the attractor solution, though that approach to equilibrium is not captured in our treatment.  Finally, we also note that our treatment assumes that the gauge group begins unconfined at the initial temperature, i.e. $\alpha(T_0)<1$.  This condition is determined by the running of $\alpha$ with temperature, and thus depends non-trivially on the choices of $N_c$ and $\alpha(T_\text{eq})$.  For $N_c=3$ and $\alpha(T_\text{eq})=0.1$, it amounts to $T_0\gtrsim T_\text{eq}/170$.

\section{Inflation begins in a vacuum}
\label{app:vac}

Here, we derive in detail the conditions presented in Sec.~\ref{sec:vac}. Recall that the exponential production of gauge modes during inflation arises due to a tachyonic instability in the gauge equations of motion~\cite{Anber:2009ua}. Let us define the gauge field operator\footnote{Throughout this appendix, we will use $A$ to refer to the RMS value of the operator $\vec{\hat A}^a$ for a single color.  We use $A_\pm(k,t)$ to refer to the mode function at a particular momentum $k$ (again for a single color).  Note that $A$ has dimension 1, while $A_\pm(k,t)$ has dimension $-1/2$.} $A$ in the Lorenz gauge with mode expansion
\be
A = \sum_{\lambda=\pm}\int \frac{d^3 \vec{k}}{(2\pi)^3}\left[A_{\lambda}(t,\vec{k})\vec{\epsilon}_{\lambda}(\vec{k})a_{\lambda}^k e^{i\vec{k}\cdot\vec{x}} + \mathrm{h.c.}\right]
\ee
where $A_{\pm}$ are the two helicity modes, $a_{\lambda}^k$ and $(a_{\lambda}^k)^\dagger$ are the creation/annihilation operators, and the helicity operators $\vec{\epsilon}_{\pm}(\vec{k})$ are defined such that $\vec{k}\cdot \vec{\epsilon}_{\pm}(\vec{k}) = 0$ and $\vec{k}\times \vec{\epsilon}_{\pm}(\vec{k}) = \mp i |\vec{k}| \vec{\epsilon}_{\pm}(\vec{k})$.

We will initially neglect the expansion of the universe, nonlinear terms arising from the non-Abelian nature of the $SU(3)$ gauge fields, and backreaction on the inflaton (assumptions that will be reevaluated later), and derive the equations of motion for the modes:
\begin{equation}\label{eom}
\ddot A_\pm-k\left(\frac{\alpha\dot\phi}{2\pi f}\mp k\right)A_\pm=0.
\end{equation}  We focus on the $+$ polarization in the case where $\dot\phi>0$.  It is clear from the equation of motion that modes with $k\leq k_\text{max}\equiv\frac{\alpha\dot\phi}{2\pi f}$ exhibit a tachyonic instability. Specifically, they grow as
\begin{equation}
A_+(k,t)\propto e^{\beta t},~~~~~\beta=\sqrt{k\cdot(k_\text{max}-k)}.
\end{equation}
Note that $\beta$ is maximized for $k=\frac{k_\text{max}}2$.  Thus most modes which are produced by this mechanism will be produced with momentum near $\frac{k_\text{max}}2$.  Since we are primarily interested in subhorizon modes, we will assume $k_\text{max}\gg2H$ (which will turn out to be weaker than the other conditions we derive). 

We wish to determine under what conditions the energy density in gauge modes becomes sufficiently large such that if thermalization occurs when nonlinear terms take over, the resulting bath would be sufficiently hot for the arguments of Sec.~\ref{sec:thermbath} to apply. We will therefore determine the field value $A_\text{NL}$ at which nonlinear terms become relevant, and ensure that this value is reached before Hubble expansion and backreaction effects become relevant.  Moreover, we will ensure that the energy density associated with $A_\text{NL}$ corresponds to a bath with $\Delta>H$.  We will thus check the three conditions presented in Sec.~\ref{sec:vac}, however we will state them in terms of values of $A$:
\begin{enumerate}
    \item Is $A_\text{NL}$ reached within a Hubble time?
    \item Is $A_\text{NL}\ll A_\text{slow}$, the field value at which the exponential production backreacts on the inflaton's motion?
    \item Is $\rho_R(A_\text{NL}) > \frac{\pi^2g_*}{30}(\frac H{10N_c^2\alpha^2})^4$?
\end{enumerate}
We will show that all three are met across large parts of the relevant parameter space.

Let us first compute $A_\text{NL}$, the field value at which nonlinear terms become relevant in the gauge field equations of motion.  The linear terms that appear in Eq.~\ref{eom} are of order $k^2A$.  The nonlinear terms we have neglected, on the other hand, are of order $N_cgA\partial A\sim N_cgkA^2$ and $N_c^2g^2A^3$.  We can thus estimate at what $A_\text{NL}$ these become comparable to the linear terms by taking $k$ to its dominant value $\frac{k_{\text{max}}}{2}$ and comparing the terms:
\bea\label{nonlinear}
\begin{split}
&k^2 A_\text{NL} \approx N_cg kA_\text{NL}^2\\
&\implies\left(\frac{k_\text{max}}2\right)^2A_\text{NL}\approx\frac{k_\text{max}}2\cdot N_cgA_\text{NL}^2\\
&\implies A_\text{NL}\approx\frac{k_\text{max}}{2N_cg}.
\end{split}
\eea
(The same result comes from comparing to $N_c^2g^2A^3$ as well.)

Now let us evaluate Condition 1, namely that $A_\text{NL}$ is reached within a Hubble time.  The initial condition we will take for our mode functions will be the Bunch-Davies vacuum, so that the mode function becomes
\begin{equation}
A_+(k,t)\sim\frac{e^{\beta t}}{\sqrt{2k}}.
\end{equation}
The operator $A$ can therefore be roughly estimated as
\begin{equation}
A\sim\sqrt{\int\frac{d^3k}{(2\pi)^3}|A_+|^2}\sim\frac1{2\pi}\sqrt{\int dk~ke^{2\beta t}}\sim\frac{k_\text{max}}{4\pi}e^\frac{k_\text{max}t}2
\end{equation}
where we have approximated the last integral by its value near $k=\frac{k_\text{max}}2$.\footnote{We have confirmed numerically that this approximation holds.} Then Condition 1 is simply
\bea\label{first}
\begin{split}
&A_\text{NL} \ll A|_{t=H^{-1}}\\
&\implies k_\text{max}\gg 2H\ln\frac{2\pi}{N_cg}\sim H
\end{split}
\eea
where we have taken $N_c=3$, and $g=1.1$ in the last approximation. Hence, as stated in Section~\ref{sec:vac}, this condition roughly reduces to ensuring subhorizon $k_\text{max}$, which was the parameter space of interest to begin with.

Now, let us check Condition 2, namely that the exponential production does not backreact on the inflaton's motion and slow its roll. We compute $A_\text{slow}$, the field value at which this occurs, by determining when the energy density $\rho_R$ of the radiation becomes comparable to the kinetic energy $\rho_{\dot{\phi}}=\frac{\dot\phi^2}2$ in the inflaton field.  Each mode $k$ with amplitude $A_+(k,t)$ will have energy equal to
\bea
\begin{split}
\frac{E^2+B^2}2&=\frac{|\dot A_+|^2+|\nabla\times A_+|^2}2\\
&=\frac{(\beta^2+k^2)|A_+|^2}2\\
&=\frac{kk_\text{max}|A_+|^2}2.
\end{split}
\eea
Then integrating over all modes and summing over all colors, we find the total energy density of the radiation to be
\begin{equation}\label{eq:Arho}
\rho_R\approx(N_c^2-1)\int\frac{d^3k}{(2\pi)^3}\frac{kk_\text{max}|A_+|^2}2\sim\frac{(N_c^2-1)k_\text{max}^2A^2}4,
\end{equation}
Then $\rho_R\sim\rho_{\dot{\phi}}$ at
\begin{equation}
A_\text{slow}\sim\frac{\sqrt2\dot\phi}{N_ck_\text{max}}
\end{equation}
Condition 2 then becomes
\bea
\label{eq:appcond3}
\begin{split}
A_\text{NL}&\ll A_\text{slow}\implies k_\text{max}\ll\sqrt[4]8\sqrt{g\dot\phi}\sim2\sqrt{\dot\phi}
\end{split}
\eea
again for $g=1.1$.

Finally let us consider Condition 3, namely that the energy density $\rho_R$ associated with $A_\text{NL}$ should be greater than $\frac{\pi^2g_*}{30}(\frac H{10N_c^2\alpha^2})^4$, so that if the sector thermalizes, it will have $\Delta>H$.  Using Eq.~\ref{eq:Arho}, we can write this condition as
\bea
\begin{split}
    &\rho_R(A_\text{NL})\gg\frac{\pi^2g_*}{30}\left(\frac H{10N_c^2\alpha^2}\right)^4\\
    &\implies k_\text{max}\gg\sqrt[4]{\frac{2\pi^3g_*}{15\alpha^7}}\frac H{5N_c^2}\sim4H,
\end{split}
\eea
where we have taken $N_c=3$, $\alpha=0.1$, and $g_*=17$.  This condition is marginally stronger than Condition 1 and the condition that the modes of interest are subhorizon.

Taking $\dot\phi$ to be the terminal velocity set by Hubble friction,\footnote{\label{ftnt:velocity}If $\dot\phi$ begins at a lower velocity, it will speed up and reach a terminal velocity within a Hubble time.  If this terminal velocity is the one set by Hubble friction, then the above treatment is applicable.  It is however possible that it reaches a lower terminal velocity set by the production of gauge modes~\cite{Anber:2009ua}.  Such a terminal velocity has an associated steady-state occupation of gauge modes $A_\text{terminal}$.  Because such a steady state requires the gauge modes to drain most of the kinetic energy of the inflaton, the energy density of $A_\text{terminal}$ is by definition comparable to that of $A_\text{slow}$ (computed with the $\dot\phi$ set by Hubble friction).  It follows that $A_\text{terminal}\sim A_\text{slow}$, and so if Eq.~\ref{eq:appcond3} is satisfied (again with the $\dot\phi$ set by Hubble friction), then nonlinearities will take over before reaching this steady state.  Thus Eqs.~\ref{eq:condition1}--\ref{eq:condition3} guarantee the production of a thermal bath for \emph{any} initial velocity.} we can rewrite these limits in terms of $f$ and $\epsilon_V$. We have
\begin{equation}
\dot\phi=\frac{V'}{3H}=\frac{\sqrt{2\epsilon_V}V}{3HM_\text{pl}}=\sqrt{2\epsilon_V}HM_\text{pl},
\end{equation}
so
\begin{equation}
\frac{k_\text{max}}H=\frac{\alpha\sqrt{2\epsilon_V}}{2\pi}\cdot\frac{M_\text{pl}}f
\end{equation}
hence we can substitute these expressions into our limits on $k_{\text{max}}/H$ to express our conditions respectively as
\begin{align}
\label{eq:condition1}
\text{Condition 1:}~~~~\frac{f}{M_{\text{pl}}}&\ll\frac{\alpha\sqrt{2\epsilon_V}}{4\pi\ln\frac{2\pi}{N_cg}}\\
\label{eq:condition2}
\text{Condition 2:}~~~~\frac{f}{M_{\text{pl}}}&\gg\sqrt[4]{\frac{\alpha^3\epsilon_V}{256\pi^5}}\sqrt{\frac H{M_\text{pl}}}\\
\label{eq:condition3}
\text{Condition 3:}~~~~\frac{f}{M_{\text{pl}}}&\ll5N_c^2\sqrt[4]{\frac{15\alpha^{11}\epsilon_V^2}{8\pi^7g_*}}.
\end{align}

%\textbf{Note:} In the original version of this paper, the condition for thermalization was computed using the 3-to-2 gluon scattering rate. This is, however, not applicable to the produced gauge modes because while they are produced with $k^2 > 0$, they obey tachyonic dispersion relations with $\omega^2 < 0$, and hence a naive application of on-shell particle scattering rates is not justified. In light of this, this version of the paper instead solely computes when the energy density in these gauge modes becomes large enough that \textit{if} thermalization occurs (as might be expected by entropic arguments when non-linearities in the equation of motion become relevant), the resulting bath has a high enough temperature for the arguments of Sec.~\ref{sec:thermbath} to apply.  We thank Junwu Huang for bringing this subtlety to our attention and for several useful discussions on this point.

\newpage

\end{appendix}

\bibliography{ref}

\end{document}